\documentclass[prb,twocolumn,amsmath,amssymb]{revtex4}

\usepackage{graphicx}
\usepackage{dcolumn}
\usepackage{bm}

\begin{document}


\title{Ground-State Energy and Condensate Density of a Dilute Bose Gas Revisited}

\author{Kazumasa T{\sc sutsui} and Takafumi K{\sc ita}}
\affiliation{Department of Physics, Hokkaido University, Sapporo 060-0810, Japan}%

\date{\today}

\begin{abstract}
The ground-state energy per particle $E/N$ and condensate density $n_0$ of a dilute Bose gas are studied with 
a self-consistent perturbation expansion satisfying the Hugenholtz-Pines theorem and conservation laws simultaneously.
A class of Feynman diagrams for the self-energy, which has escaped consideration so far, 
is shown to add an extra constant $c_{{\rm ip}}\sim O(1)$
to the expressions reported by Lee, Huang, and Yang [Phys.\ Rev.\ {\bf  106}
(1957) 1135] as $E/N= (2\pi\hbar^2 a n/m) \bigl[1+(128/15\sqrt{\pi}+16c_{{\rm ip}}/5) \sqrt{a^3n}\, \bigr]$  and ${n_0}/{n}=1-(8/3\sqrt{\pi}+c_{{\rm ip}}) \sqrt{a^3n}$, 
where $a$, $n$, and $m$ are are the $s$-wave scattering length, particle density, and particle mass, respectively.
We present a couple of estimates for $c_{\rm ip}$; the third-order perturbation expansion  yields $c_{{\rm ip}}=0.412$. 
\end{abstract}

\maketitle

Lee {\em et al}.\ \cite{LHY57} studied the ground-state properties of a single-component homogeneous dilute Bose gas
based essentially on Bogoliubov theory\cite{Bogoliubov47} at zero temperature.
They obtained widely accepted expressions for  the energy per particle and condensate density as\cite{LHY57,AGD63,FW71}
\begin{subequations}
\label{LHY}
\begin{equation}
\frac{E}{N}= \frac{2\pi\hbar^2 a n}{m} \left(1+\frac{128}{15\sqrt{\pi}} \sqrt{a^3n}\right),
\label{LHY1}
\end{equation}
\begin{equation}
\frac{n_0}{n}= 1-\frac{8}{3\sqrt{\pi}} \sqrt{a^3n},
\label{LHY2}
\end{equation}
\end{subequations}
where $a$, $n$, and $m$ are are the $s$-wave scattering length, particle density, and particle mass, respectively.
On the other hand, a standard finite-temperature extension of Bogoliubov theory is known to yield
an unphysical energy gap in the single-particle spectrum\cite{GA59,Griffin96} 
in contradiction to the Hugenholtz-Pines theorem.\cite{HP59}
This issue was discussed extensively by Hohenberg and Martin\cite{HM65} in connection with conservation laws.

Recently, a self-consistent perturbation expansion has been formulated  for condensed Bose systems so that the
Hugenholtz-Pines theorem and conservation laws are satisfied simultaneously.\cite{Kita09,Kita10}
A notable prediction of this theory\cite{Kita11} is that there is a new class of Feynman diagrams for the self-energy,
i.e., those that may be classified as {\em improper} or {\em reducible} following the conventional terminology
and should certainly be excluded from its definition in the normal state.\cite{LW60,AGD63,FW71}
The rationale for their existence in condensed Bose systems is that they are indispensable for three exact statements, i.e., the Hugenholtz-Pines theorem, conservation laws, and an
identity for the interaction energy, to be satisfied order by order simultaneously in the self-consistent perturbation expansion.\cite{Kita09}
 Thus, the conventional procedure of constructing the self-energy in the Dyson-Beliaev equation with the concept of {\em properness} (or {\em irreducibility})\cite{AGD63,FW71} cannot be justified.
Since they have been overlooked, it will be worth looking at how the {\em improper} diagrams affect the standard results for the dilute Bose gas.\cite{AGD63,FW71}

In the present paper, we will show that the {\em improper} diagrams change eq.\ (\ref{LHY}) into
\begin{subequations}
\label{LHY-a}
\begin{equation}
\frac{E}{N}=  \frac{2\pi\hbar^2 a n}{m} \left[1+\frac{16}{5}\left(\frac{8}{3\sqrt{\pi}}+c_{\rm ip} \right)\sqrt{a^3n}\right],
\label{LHY-a1}
\end{equation}
\begin{equation}
\frac{n_0}{n}= 1-\left(\frac{8}{3\sqrt{\pi}}+c_{\rm ip}\right) \sqrt{a^3n}\,,
\label{LHY-a2}
\end{equation}
\end{subequations}
where $c_{{\rm ip}}$ is a numerical constant of order $1$.
We will subsequently estimate $c_{{\rm ip}}$ using a couple of approximations.
One of them, i.e., the perturbation expansion up to the third order, will be shown to yield $c_{{\rm ip}}=0.412$.

We consider a homogeneous system of identical Bose particles with mass $m$ and spin $0$ interacting via
the contact potential $U\delta({\bf r}_{1}-{\bf r}_{2})$.
The Hamiltonian is given by
\begin{equation}
H = \sum_{\bm p} (\epsilon_p -\mu)c_{\bm p}^{\dagger} c_{\bm p} + \frac{U}{2V}\sum_{{\bm p}_1{\bm p}_2{\bm q}}c_{{\bm p}_1+{\bm q}}^{\dagger} 
c_{{\bm p}_2-{\bm q}}^{\dagger} c_{{\bm p}_2}c_{{\bm p}_1} ,
\end{equation}
where ${\bm p}$, $\epsilon_p$, $\mu$, and $V$ are the momentum, kinetic energy,  chemical potential, and volume,
and $c_{\bm p}^{\dagger}$ and $c_{\bm p}$ are the creation and annihilation operators, respectively. 
It is convenient to set
$\hbar=2m=k_{\rm B}=T_{\rm c}^{0}=1$,
where $k_{\rm B}$ denotes the Boltzmann constant and $T_{\rm c}^{0}$ is the transition temperature of ideal Bose-Einstein condensation.\cite{FW71}
Thus, the kinetic energy is expressed  simply as $\epsilon_p=p^2$.
Ultraviolet divergences inherent in the continuum model are removed here
by introducing a momentum cutoff $p_{\rm c}\gg 1$. 
However, our final results will be free from $p_{\rm c}$, as seen below.
It is standard in the low-density limit to remove $U$ in favor of the $s$-wave scattering length $a$.
They are connected in the conventional units by
$$
\frac{m}{4\pi \hbar^2 a}= \frac{1}{U}+\int\frac{{\rm d}^3 p}{(2\pi\hbar)^3} \frac{\theta (p_{\rm c}-p)}{2\epsilon_p } ,
$$
with $\theta (x)$ the step function, which in the present units reads
$1/8\pi a =1/U+p_{\rm c}/4\pi^2$. 
We will focus on the limit $a \rightarrow  0$ and choose $p_{\rm c}$ so that $1\ll p_{\rm c}\ll \pi/2a$ is satisfied.
Thus, we can set
\begin{equation}
U=\frac{8\pi a}{1-(2/\pi) p_{\rm c}a}\approx 8\pi a \left(1+\frac{2}{\pi}p_{\rm c}a\right)
\label{U-a}
\end{equation}
as an excellent approximation for the present purpose.

Let us recapitulate some relevant results from ref.\ \onlinecite{Kita09}.
Green's function for the condensed Bose system can be expressed in the Nambu representation as
\begin{subequations}
\label{Green'sFn}
\begin{equation}
\hat{G}_{\vec{p}}
\equiv
\begin{bmatrix}
\vspace{1mm}
G_{\vec{p}} & F_{\vec{p}} \\
-\bar{F}_{\vec{p}} & -\bar{G}_{\vec{p}}
\end{bmatrix} ,
\label{hatG_p}
\end{equation}
where $\vec{p}=({\bm p},i\varepsilon_\ell)$ with $\varepsilon_\ell\equiv
2\ell\pi T$ ($\ell=0,\pm 1,\pm 2,\cdots$) the Matsubara frequency.
The upper elements satisfy $G_{\vec{p}}=G_{\vec{p}^{\,*}}^*$ and
$F_{\vec{p}}=F_{-\vec{p}}$, and a barred quantity generally denotes
$\bar{G}_{\vec{p}}\equiv G_{-\vec{p}^{\,*}}^{*}$.
The matrix Green's function obeys the Dyson-Beliaev equation
\begin{equation}
\hat{G}_{\vec{p}}
=
\begin{bmatrix}
\vspace{1mm}
i\varepsilon_\ell-\epsilon_p -\Sigma_{\vec{p}}+\mu & -\Delta_{\vec{p}}
\\
\bar{\Delta}_{\vec{p}} & i\varepsilon_\ell+\epsilon_p +\bar{\Sigma}_{\vec{p}}-\mu 
\end{bmatrix}^{-1},
\label{DB}
\end{equation}
\end{subequations}
which may also be regarded as defining the self-energies
$\Sigma_{\vec{p}}$ and $\Delta_{\vec p}$.
In the self-consistent perturbation expansion, 
they are obtained from a functional $\Phi[\hat{G}_{\vec{p}},n_0]$ as
\begin{subequations}
\label{Phi-deriv}
\begin{equation}
\Sigma_{\vec{p}}=-T^{-1}\frac{\delta\Phi}{\delta G_{\vec p}}\,,
\hspace{10mm}
\Delta_{\vec{p}}=2T^{-1}\frac{\delta\Phi}{\delta \bar F_{\vec p}}\, .
\label{Sigma-Phi}
\end{equation}
In addition, $\Phi$ satisfies 
\begin{equation}
\frac{1}{V} \frac{\delta \Phi}{\delta n_0}=\Sigma_{\vec{0}}-\Delta_{\vec{0}} \,.
\label{n_0-Phi}
\end{equation}
\end{subequations}
The substitution of eq.\ (\ref{Sigma-Phi}) into eq.\ (\ref{DB}) yields self-consistent (i.e., nonlinear) equations 
for $G_{\vec p}$ and $F_{\vec p}$.
It also follows from eq.\ (\ref{n_0-Phi}) that the stationarity condition $\delta \Omega/\delta n_0 =0$ 
for the thermodynamic potential $\Omega$ is equivalent to the Hugenholtz-Pines relation 
\begin{equation}
\mu=\Sigma_{\vec{0}}-\Delta_{\vec{0}} .
\label{HP}
\end{equation}
These are exact statements. 
It has been shown that the key functional $\Phi$ can be constructed as a power-series expansion in $U$
in such a way that eq.\ (\ref{HP}), conservation laws, and an exact relation for the interaction energy are fulfilled simultaneously 
order by order.

\begin{figure}[b]
        \begin{center}
                \includegraphics[width=0.8\linewidth]{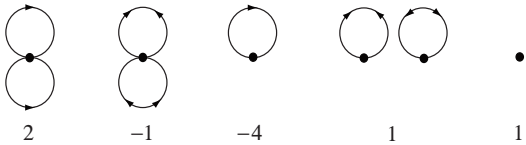}
        \end{center}
        \vspace{-4mm}
        \caption{Feynman diagrams for $\Phi^{(1)}$. 
         A filled circle denotes $2U$, a line with an arrow (two arrows) represents $G$
         (either $F$ or $\bar{F}$) in eq.\ (\ref{hatG_p}) as in the theory of superconductivity,\cite{AGD63,FW71}
         and every missing line in the last three diagrams corresponds to $n_0$.
         The number below each diagram indicates its relative weight,
         which should be multiplied by $1/4$ to obtain the absolute weight.\label{Fig1}}
\end{figure}

The first-order functional $\Phi^{(1)}$ is given graphically in Fig.\ \ref{Fig1}, 
which can be shown to reproduce eq.\ (\ref{LHY}) as follows.
The differentiations of eq.\ (\ref{Sigma-Phi}) correspond to removing a line of 
$G_{\vec p}$ and $\bar F_{\vec p}$, respectively, from every diagram in Fig.\ \ref{Fig1} in all possible ways.
Hence, $\Sigma^{(1)}$ and $\Delta^{(1)}$ are obtained as
\begin{subequations}
\label{SigmaDelta^(1)}
\begin{equation}
\Sigma^{(1)}=2U n ,
\label{Sigma^(1)}
\end{equation}
\begin{equation}
\Delta^{(1)}=U \biggl( n_{0}  -\sum_{\vec{p}}F_{\vec{p}}^{(1)}\biggr) ,
\label{Delta^(1)}
\end{equation}
where
\begin{equation}
n = n_{0}  -\sum_{\vec{p}}G_{\vec{p}}^{(1)}{\rm e}^{i\varepsilon_\ell 0_+}
\label{n^(1)}
\end{equation}
is the particle density with $0_+$ an infinitesimal positive constant,\cite{LW60,AGD63} 
and the summation over $\vec p$ denotes 
$$
\sum_{\vec{p}}\equiv T\sum_{\ell=-\infty}^\infty \int\frac{{\rm d}^3p}{(2\pi)^3} \,\,\stackrel{T\rightarrow 0}{\longrightarrow}
\,\, \int_{-\infty}^{\infty} \frac{{\rm d}\varepsilon_\ell}{2\pi}\int\frac{{\rm d}^3p}{(2\pi)^3} .
$$
Thus, the first-order self-energies $\Sigma^{(1)}$ and $\Delta^{(1)}$ have no $\vec{p}$ dependence,
and eq.\ (\ref{HP}) reduces to
\begin{equation}
\mu^{(1)}=\Sigma^{(1)}-\Delta^{(1)} .
\label{HP^(1)}
\end{equation}
\end{subequations}
Substituting eqs.\ (\ref{Sigma^(1)}), (\ref{Delta^(1)}), and (\ref{HP^(1)}) into eq.\ (\ref{DB}) and carrying out matrix inversion, 
we obtain the upper elements of $\hat G_{\vec p}$ in eq.\ (\ref{hatG_p}) as
\begin{equation}
\begin{bmatrix}
\vspace{2mm}
G_{\vec p}^{(1)}\\
F_{\vec p}^{(1)}
\end{bmatrix}
=\frac{-1}{\varepsilon_\ell^2+\epsilon_p (\epsilon_p +2\Delta^{(1)})}
\begin{bmatrix}
\vspace{2mm}
i\varepsilon_\ell+\epsilon_p +\Delta^{(1)} \\
\Delta^{(1)}
\end{bmatrix} .
\label{GF^(1)}
\end{equation}
With eq.\ (\ref{GF^(1)}), we can perform the summations over $\vec{p}$ in eqs.\ (\ref{Delta^(1)}) and (\ref{n^(1)}) analytically
to obtain
\begin{subequations}
\label{Delta^(1)-n-2}
\begin{equation}
\Delta^{(1)}=U\left[n_0+\frac{\Delta^{(1)}}{4\pi^2}p_{\rm c}-\frac{\bigl(2\Delta^{(1)}\bigr)^{3/2}}{8\pi^2}\right],
\label{Delta^(1)-2}
\end{equation}
\begin{equation}
n= n_0+\frac{\bigl(2\Delta^{(1)}\bigr)^{3/2}}{24\pi^2} ,
\label{n-2}
\end{equation}
\end{subequations}
where we have set $p_{\rm c}\rightarrow \infty$ in the second convergent integral;
this procedure will be adopted throughout.
The substitution of eq.\ (\ref{U-a}) into eq.\ (\ref{Sigma^(1)}) yields
\begin{subequations}
\label{SigmaDelta^(1)-3}
\begin{equation}
\Sigma^{(1)}=16\pi a n \left(1+\frac{2}{\pi}p_{\rm c}a\right).
\label{Sigma^(1)-3}
\end{equation}
For $\Delta^{(1)}$, eqs.\ (\ref{U-a})  and (\ref{Delta^(1)-n-2}) indicate that $\Delta^{(1)}\approx 8\pi an$ to the leading order.
Collecting next-order terms perturbatively with $p_{\rm c}a\ll 1$ in mind, we arrive at
\begin{equation}
\Delta^{(1)}=8\pi a n\left(1+\frac{4}{\pi}p_{\rm c}a-\frac{32}{3\sqrt{\pi}}\sqrt{a^3n}\right).
\label{Delta^(1)-3}
\end{equation}
\end{subequations}
Let us substitute eq.\ (\ref{SigmaDelta^(1)-3}) into eqs.\ (\ref{HP^(1)}) and (\ref{n-2}).
We then obtain
\begin{equation}
\mu^{(1)}=8\pi a n \left(1+\frac{32}{3\sqrt{\pi}}\sqrt{a^3n}\right)
\label{mu^(1)}
\end{equation}
and eq.\ (\ref{LHY2}), respectively, which are correct up to the next-to-the-leading order and also free from $p_{\rm c}$.
Finally integrating the thermodynamic relation
$\mu=\partial E/\partial N$ over the particle number $N$, we arrive at eq.\ (\ref{LHY1}) in units of $\hbar=2m=1$. 
Thus, the Lee-Huang-Yang results of eq.\ (\ref{LHY}) have been reproduced in our mean-field approximation.

\begin{figure}[t]
        \begin{center}
                \includegraphics[width=0.9\linewidth]{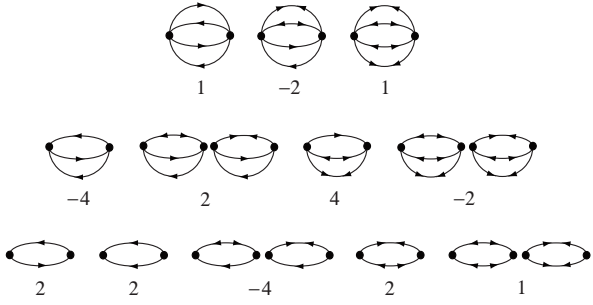}
        \end{center}
        \vspace{-4mm}
        \caption{Feynman diagrams for $\Phi^{(2)}$. 
         A filled circle denotes $2U$, a line with an arrow (two arrows) represents $G$
         (either $F$ or $\bar{F}$) in eq.\ (\ref{hatG_p}),
         and every missing line in the second and third rows corresponds to $n_0$.
         The number below each diagram indicates its relative weight,
         which should be multiplied by $-1/8$ to obtain the absolute weight.\label{Fig2}}
\end{figure}

\begin{figure}[t]
        \begin{center}
                \includegraphics[width=0.8\linewidth]{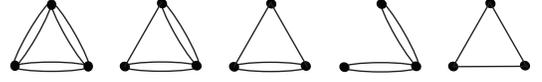}
        \end{center}
        \caption{Feynman diagrams for $\Phi^{(3)}$ drawn without arrows.}
        \label{Fig3}
\end{figure}

Now, we proceed to look into higher-order terms.
The second-order $\Phi^{(2)}$ can also be constructed uniquely so as
to satisfy eq.\ (\ref{HP}), conservation laws, and an exact relation for the 
interaction energy.\cite{Kita09}
The results are shown graphically in Fig.\ \ref{Fig2}.
For the dilute Bose gas at $T=0$, in general, the relative importance of each diagram
decreases as the number of Green's function lines in it increases, 
as our previous analysis on eq.\ (\ref{SigmaDelta^(1)}) indicates. 
Hence, the dominant contribution of the second order originates from the third-row diagrams in Fig.\ \ref{Fig2}, 
which also bring a novel structure to the self-energies.
Indeed, the differentiations of eq.\ (\ref{Sigma-Phi}) for these diagrams yield
\begin{subequations}
\label{Sigma^(23ip)}
\begin{equation}
\Sigma_{\vec p}^{(2{\rm ip})}=\Delta_{\vec p}^{(2{\rm ip})}=2(Un_0)^2(G_{\vec p}+\bar G_{{\vec p}}-F_{\vec p}-\bar F_{\vec p}) ,
\label{Sigma^(2ip)}
\end{equation}
which is classified as {\em improper} or {\em reducible} in the normal state\cite{LW60,AGD63,FW71}
and hence has been overlooked so far.
This structure contains nothing that contradicts eq.\ (\ref{DB}), which defines self-energies, however.
It is a natural consequence of the requirement that the Hugenholtz-Pines relation (\ref{HP}), which is an exact statement,
is obeyed order by order in the self-consistent perturbation expansion.
We will show shortly that it brings an additional term to eq.\ (\ref{LHY}). 
On the other hand, the diagrams in the first and second rows in Fig.\ \ref{Fig2},
which yield conventional {\em proper} or {\em irreducible} self-energies,
are at least one order of magnitude smaller at $T=0$; thus, they can be neglected for the present purpose.

The {\em improper} structure extends beyond the second order. 
In the third order, for example, the rightmost diagram in Fig.\ \ref{Fig3} also yields\cite{Kita09,Kita11}
\begin{equation}
\Sigma_{\vec p}^{(3{\rm ip})}=\Delta_{\vec p}^{(3{\rm ip})}=\frac{5}{4}(2Un_0)^3(G_{\vec p}+\bar G_{{\vec p}}-F_{\vec p}-\bar F_{\vec p})^2 ,
\label{Sigma^(3ip)}
\end{equation}
\end{subequations}
where the factor $5/4$ is the sum of $1$ and $1/2^2$ originating from the particle-hole and particle-particle bubble diagrams, respectively.
Equation (\ref{Sigma^(23ip)}) reveals common features of the leading-order {\em improper} contribution at each order: (i) they are functionals of 
\begin{equation}
f_{\vec p}\equiv G_{\vec p}+\bar G_{{\vec p}}-F_{\vec p}-\bar F_{\vec p},
\label{f_p}
\end{equation}
and (ii) the diagonal and off-diagonal self-energies are the same.
Note the symmetry $f_{\vec p}=\bar f_{\vec{p}}$.

Adding the series of eq.\ (\ref{Sigma^(23ip)}) to eqs.\ (\ref{Sigma^(1)}) and (\ref{Delta^(1)}), 
we can express the diagonal and off-diagonal self-energies as
\begin{equation}
\Sigma_{\vec p} = \Sigma^{(1)}+\Delta_{\vec p}^{({\rm ip})},\hspace{10mm}\Delta_{\vec p} = \Delta^{(1)}+\Delta_{\vec p}^{({\rm ip})},
\label{SigmaDelta-ip}
\end{equation}
respectively,
where $\Delta_{\vec p}^{({\rm ip})}\equiv \Delta_{\vec p}^{(2{\rm ip})}+\Delta_{\vec p}^{(3{\rm ip})}+\cdots$ 
satisfies $\Delta_{\vec p}^{({\rm ip})}=\bar\Delta_{{\vec p}}^{({\rm ip})}$.
The corresponding Green's functions are obtained by using eqs.\ (\ref{Green'sFn}) and (\ref{HP}) as
\begin{equation}
\begin{bmatrix}
\vspace{2mm}
G_{\vec p}\\
F_{\vec p}
\end{bmatrix}
=\frac{-1}{\varepsilon_\ell^2+\epsilon_p (\epsilon_p +2\Delta_{\vec p})}
\begin{bmatrix}
\vspace{2mm}
i\varepsilon_\ell+\epsilon_p +\Delta_{\vec p} \\
\Delta_{\vec p}
\end{bmatrix} ,
\label{GF}
\end{equation}
with
\begin{subequations}
\label{munD-new}
\begin{equation}
\mu = \Sigma^{(1)}-\Delta^{(1)} .
\label{mu-new}
\end{equation}
This expression for $\mu$ appears to be the same as eq.\ (\ref{HP^(1)}), 
but the right-hand side now should be calculated 
by using eq.\ (\ref{GF}) instead of eq.\ (\ref{GF^(1)}) in the self-consistent approximation scheme.
To be specific, $\Sigma^{(1)}$ is still given by eq.\ (\ref{Sigma^(1)}) 
but with $G_{\vec p}^{(1)}\rightarrow G_{\vec p}$ in eq.\ (\ref{n^(1)}), i.e.,
\begin{equation}
n = n_0-\sum_{\vec p}G_{\vec p}^{(1)}{\rm e}^{i\varepsilon_\ell 0_+}-\sum_{\vec p}(G_{\vec p}-G_{\vec p}^{(1)}) .
\label{n-new}
\end{equation}
This redundant expression enables us to treat the third term on the right-hand side as a correction
to the mean-field result (\ref{LHY2}).
Similarly, eq.\ (\ref{Delta^(1)}) is replaced by
\begin{equation}
\Delta^{(1)} = U\biggl[n_0-\sum_{\vec p}F_{\vec p}^{(1)}-\sum_{\vec p}(F_{\vec p}-F_{\vec p}^{(1)})\biggr].
\label{D-new}
\end{equation}
\end{subequations}
On the other hand, the substitution of eq.\ (\ref{GF}) into the series of eq.\ (\ref{Sigma^(23ip)}) yields the
equation for 
$\Delta_{\vec p}^{({\rm ip})}\equiv \Delta_{\vec p}^{(2{\rm ip})}+\Delta_{\vec p}^{(3{\rm ip})}+\cdots$ as
\begin{equation}
\Delta_{\vec p}^{({\rm ip})} = Un_0\left[
-\frac{2Un_0}{\eta_{\vec p}+\Delta_{\vec p}^{({\rm ip})}}+ \frac{10(Un_0)^2}{\bigl(\eta_{\vec p}+\Delta_{\vec p}^{({\rm ip})}\bigr)^2}
+\cdots \right],
\label{Delta^ip-eq}
\end{equation}
with
\begin{equation}
\eta_{\vec p}\equiv \frac{\varepsilon_\ell^2+\epsilon_p(\epsilon_p+2\Delta^{(1)})}{2\epsilon_p}.
\label{eta-def}
\end{equation}
Equation (\ref{Delta^ip-eq}) forms an algebraic equation for $\Delta^{({\rm ip})}$,
and we see immediately that $\Delta_{\vec p}^{({\rm ip})}\sim Un_0\sim Un$, i.e.,
it is of the same order as $\Delta^{(1)}$ in eq.\ (\ref{Delta^(1)}).

Let us transform the third term on the right-hand side of eq.\ (\ref{n-new}) by
using eqs.\ (\ref{GF^(1)}) and  (\ref{GF}), noting the summation over $\vec p$ below eq.\ (\ref{n^(1)}) at $T=0$,
and making a change of variables as
\begin{equation}
\epsilon_p=Un \tilde \epsilon_p,\hspace{5mm}\varepsilon_\ell=Un \tilde \varepsilon_\ell,\hspace{5mm}
\Delta_{\vec p}=Un \tilde \Delta_{\vec p}.
\label{tilde-def}
\end{equation}
The result may be written as
\begin{equation}
\sum_{\vec{p}} \left(G_{\vec{p}}-G_{\vec{p}}^{(1)}\right) = -\left(\frac{Un}{8\pi}\right)^{\!\! 3/2}c_{\rm ip} ,
\label{DeltaG-c_G}
\end{equation}
where $c_{\rm ip} $ is given by
\begin{eqnarray}
&&\hspace{-12mm}
c_{\rm ip} \equiv  \frac{4\sqrt{2}}{\pi^{3/2}}\int_0^\infty {\rm d}\tilde \epsilon_p \tilde \epsilon_p^{1/2}  \int_0^\infty {\rm d}\tilde \varepsilon_\ell
\nonumber \\
&&\hspace{-3.5mm}
\times \frac{(\tilde \varepsilon_\ell ^2-\tilde \epsilon_p^2 )\tilde \Delta_{\vec p}^{({\rm ip})}}{[\tilde \varepsilon_\ell ^2
+\tilde \epsilon_p(\tilde \epsilon_p+2\tilde \Delta_{\vec p})][\tilde \varepsilon_\ell ^2+\tilde \epsilon_p(\tilde \epsilon_p+2\tilde \Delta^{(1)})]} .
\label{c_G-def}
\end{eqnarray}
When substituting eq.\ (\ref{DeltaG-c_G}) into eq.\ (\ref{n-new}) for the present purpose, 
we can replace $(Un/8\pi)^{3/2}$ by the leading-order expression $(na)^{3/2}$ with $\tilde \Delta^{(1)}\rightarrow 1$
in eq.\ (\ref{c_G-def}); see eqs.\ (\ref{U-a}), (\ref{SigmaDelta^(1)}), and (\ref{Delta^(1)-n-2}) on this point. 
Equation (\ref{Delta^ip-eq}) is also approximated with $Un_0\approx Un$ as
\begin{equation}
\tilde\Delta_{\vec p}^{({\rm ip})} =
-\frac{2}{\tilde\Delta_{\vec p}^{({\rm ip})}+\tilde\eta_{\vec p}}+ \frac{10}{\bigl(\tilde\Delta_{\vec p}^{({\rm ip})}+\tilde\eta_{\vec p}\bigr)^2}
+\cdots ,
\label{x_p-eq}
\end{equation}
where $\tilde\eta_{\vec p}$ is defined by
\begin{equation}
\tilde\eta_{\vec p}\equiv (\tilde \varepsilon_\ell^2+\tilde \epsilon_p^2)/2\tilde \epsilon_p+1 .
\label{a_p-def}
\end{equation}
Equation (\ref{x_p-eq}) with eq.\ (\ref{a_p-def}) implies that $\tilde\Delta_{\vec p}^{({\rm ip})}$ 
is a function of only $x\equiv \tilde \varepsilon_\ell^2/\tilde \epsilon_p+\tilde \epsilon_p$.
Hence, it is convenient to make a further change of variables in eq.\ (\ref{c_G-def}):
\begin{equation}
x\equiv \frac{\tilde \varepsilon_\ell^2}{\tilde \epsilon_p}+\tilde \epsilon_p,\hspace{3mm}
\xi\equiv \tilde \epsilon_p ,\hspace{3mm}b(x)\equiv \frac{x}{2}+1,\hspace{3mm}y(x)\equiv \tilde\Delta_{\vec p}^{({\rm ip})},
\label{CV1}
\end{equation}
to transform it to
\begin{eqnarray}
&&\hspace{-10mm}
c_{\rm ip} 
=\frac{2\sqrt{2}}{\pi^{3/2}}
\int_0^\infty \!{\rm d}\xi    \int_\xi^\infty \!{\rm d}x
 \frac{(x-2\xi )y(x)}{\sqrt{x-\xi }[x+2+2y(x)](x+2)}
\nonumber \\
&&\hspace{-4.8mm}
=-\frac{4\sqrt{2}}{3\pi^{3/2}}
\int_0^\infty {\rm d}x   \frac{x^{3/2} y(x)}{[x+2+2y(x)](x+2)}.
\label{c_G-def2}
\end{eqnarray}
The last expression has been obtained 
by changing the order of integrations
and subsequently integrating over $\xi$.
A similar analysis on the third term in the square brackets of eq.\ (\ref{D-new}) yields
\begin{equation}
\sum_{\vec{p}} \left(F_{\vec{p}}-F_{\vec{p}}^{(1)}\right) = \left(\frac{Un}{8\pi}\right)^{\!\! 3/2} c_F ,
\label{DeltaF-c_F}
\end{equation}
where $c_F$ is the same as eq.\ (\ref{c_G-def}) except for
$\tilde \varepsilon_\ell ^2-\tilde \epsilon_p^2\rightarrow
-\tilde \varepsilon_\ell ^2-\tilde \epsilon_p^2$
in the numerator. 
Transforming it in the same manner as from eq.\  (\ref{c_G-def}) into eq.\ (\ref{c_G-def2}),
we find that
\begin{equation}
c_F =3c_{\rm ip} .
\label{c_F-def}
\end{equation}
Let us substitute eqs.\ (\ref{DeltaG-c_G}) and (\ref{DeltaF-c_F}) with eq.\ (\ref{c_F-def}) into eq.\ (\ref{munD-new}).
We then find that the {\em improper} contribution changes eqs.\ (\ref{LHY2}) and (\ref{mu^(1)}) into eqs.\ (\ref{LHY-a2}) and
\begin{equation}
\mu=8\pi a n \left[1+4\left(\frac{8}{3\sqrt{\pi}}+c_{\rm ip}\right)\sqrt{a^3n}\right] ,
\label{mu-new2}
\end{equation}
respectively.
Finally using the thermodynamic relation
$\mu=\partial E/\partial N$ to integrate eq.\ (\ref{mu-new2}) over $N$, we arrive at eq.\ (\ref{LHY-a1}). 
Equation (\ref{LHY-a}) with eq.\ (\ref{c_G-def2}) are correct up to the next-to-the-leading order.

Finally, let us estimate $c_{\rm ip}$ by using two different approximations.
First, retaining only the first two terms on the right-hand side of eq.\ (\ref{x_p-eq}) yields
a cubic equation for $y=y(x)$, which is defined in eq.\ (\ref{CV1}), as
\begin{equation}
y^3-2by^2+(b^2+2)y-2(b+5)=0.
\label{x_p-eq-3rd}
\end{equation}
It has a single real solution $y=y(x)$ for $0\leq x\leq \infty$
that approaches $0$ continuously as $x\rightarrow\infty$, 
in accordance with $\Delta^{({\rm ip})}(i\varepsilon_\ell)\rightarrow 0$ as $|\varepsilon_\ell |\rightarrow \infty$.
Substituting it  into eq.\ (\ref{c_G-def2}) and performing the integration,
we obtain $c_{\rm ip}=0.412$.  Hence, the factor $128/15\sqrt{\pi}=4.81$ in eq.\ (\ref{LHY1}) is changed into $(16/5)(8/3\sqrt{\pi}+c_{\rm ip})=6.13$ in eq.\ (\ref{LHY-a1}).
Second, eq.\ (\ref{Sigma^(3ip)}) may be generalized to
the $n$th-order {\em improper} terms ($n=3,4,\cdots$) originating from the particle-hole and particle-particle bubble diagrams
as
\begin{equation}
\Sigma^{(n{\rm ip})}_{\vec{p}}=\Delta^{(n{\rm ip})}_{\vec{p}}
=\left(1+\frac{1}{2^{n-1}}\right)(2Un_{0})^{n}f_{\vec{p}}^{n-1},
\label{Sigma^(n)-approx1}
\end{equation}
where $f_{\vec p}$ is defined by eq.\ (\ref{f_p}). 
Collecting the series for $n=3,4,\cdots,\infty$ together with eq.\ (\ref{Sigma^(2ip)}) of the second order,
we obtain
\begin{eqnarray}
&&\hspace{-10mm}
\Delta_{\vec{p}}^{({\rm ip})}=\frac{(2Un_{0})^{2}f_{\vec{p}}}{ 1-2Un_{0}f_{\vec{p}} }+\frac{2(Un_{0})^{2}f_{\vec{p}}}{1-Un_{0}f_{\vec{p}}}-4(Un_{0})^{2}f_{\vec{p}}
\nonumber \\
&&\hspace{-1.5mm}
= -\frac{(2Un_{0})^{2}}{\eta_{\vec{p}}+2Un_0 +\Delta_{\vec{p}}^{({\rm ip})}}
-\frac{2(Un_{0})^{2}}{\eta_{\vec{p}}+Un_0+\Delta_{\vec{p}}^{({\rm ip})}}
\nonumber \\
&&\hspace{2.5mm}
+\frac{4(Un_{0})^{2}}{\eta_{\vec{p}}+\Delta_{\vec{p}}^{({\rm ip})}}.
\label{tildeDelta0}
\end{eqnarray}
This self-consistent equation for $\Delta_{\vec{p}}^{({\rm ip})}$ is transformed in terms of the quantities in eq.\ (\ref{CV1}) into
\begin{eqnarray}
&&\hspace{-10mm}
y^4-3(b-1)y^3+(3b^2-6b+4)y^2
\nonumber \\
&&\hspace{-10mm}
-(b^3-3b^2+6b+4)y
+2(b^2+2b-4)=0,
\label{x_p-eq-4th}
\end{eqnarray}
which also has a single real solution $y=y(x)$ for $0\leq x\leq \infty$
that approaches $0$ continuously as $x\rightarrow\infty$.
Substituting it into eq.\ (\ref{c_G-def2}), we obtain $c_{\rm ip}=0.563$.
Thus, we may expect that $c_{\rm ip}$ is of order $1$.
The exact value of $c_{\rm ip}$ can only be reached by
collecting all the leading-order {\em improper} terms, however, 
meaning that further studies are required for a quantitative estimate of $c_{\rm ip}$.
On the other hand, a diffusion Monte Carlo study was performed on eq.\ (\ref{LHY});\cite{GBC99}
the data in Table I of ref.\ \onlinecite{GBC99} for $a^3n\leq  10^{-2}$ yield  the value $4.3(1)$, rather than $128/15\sqrt{\pi}=4.81$ from Eq.\ (\ref{LHY1}),
thus showing a clear deviation from the Lee-Huang-Yang expression beyond the numerical uncertainty.
On the other hand, an experiment on eq.\ (\ref{LHY1}) in a trap potential reported a value 4.5(7),\cite{Navon11} which still requires improvement for the present purpose, however.

\acknowledgments
We are grateful to D. Hirashima for a useful discussion on the diffusion Monte Carlo study.
This work is supported in part by JSPS Grant (C) No.\ 22540356.

\end{document}